# Advancing Email Spam Detection: Leveraging Zero-Shot Learning and Large Language Models

Ghazaleh Shirvani
Carleton University
ghazalehshirvani@cmail.carleton.ca
Saeid Ghasemshirazi
Carleton University
saeidghasemshirazi@cmail.carleton.ca

## Abstract

Email spam detection is a critical task in modern communication systems, essential for maintaining productivity, security, and user experience. Traditional machine learning and deep learning approaches, while effective in static settings, face significant limitations in adapting to evolving spam tactics, addressing class imbalance, and managing data scarcity. These challenges necessitate innovative approaches that reduce dependency on extensive labeled datasets and frequent retraining. This study investigates the effectiveness of Zero-Shot Learning using FLAN-T5, combined with advanced Natural Language Processing (NLP) techniques such as BERT for email spam detection. By employing BERT to preprocess and extract critical information from email content, and FLAN-T5 to classify emails in a Zero-Shot framework, the proposed approach aims to address the limitations of traditional spam detection systems. The integration of FLAN-T5 and BERT enables robust spam detection without relying on extensive labeled datasets or frequent retraining, making it highly adaptable to unseen spam patterns and adversarial environments. This research highlights the potential of leveraging zero-shot learning and NLPs for scalable and efficient spam detection, providing insights into their capability to address the dynamic and challenging nature of spam detection tasks.

## 1 Introduction

Spam detection remains a critical challenge in modern communication systems. Spam messages disrupt communication and reduce productivity. They also introduce security risks, such as phishing and malware attacks[9]. Traditional methods for spam detection, including machine learning and rule-based systems, face several limitations. These methods depend heavily on large, labeled datasets, which are often expensive and time-consuming to create [2]. Furthermore, they struggle to adapt to evolving spam patterns, known as concept drift. These methods are also vulnerable to adversarial attacks, where spammers intentionally alter content to evade detection [3].

Recent research has explored Zero-Shot Learning (ZSL) as a promising alternative to traditional approaches. ZSL allows models to classify unseen data by leveraging shared semantic information, making it particularly suitable for spam detection, where new spam types frequently emerge [9]. However, existing ZSL-based methods often lack the semantic depth and contextual understanding needed to handle the evolving patterns of spam content [5]. These gaps in semantic representation and adaptability have limited the widespread adoption of ZSL in real-world spam detection scenarios.

To address these challenges, this study integrates ZSL with advanced Natural Language Processing models, leveraging the complementary strengths of BERT and FLAN-T5. BERT is used to generate concise summaries of email content, helping to reduce noise and focus on key information for more effective analysis. These summaries are then classified using FLAN-T5, a large, pretrained language model specifically designed for tasks requiring generalization and adaptability. By embedding both the summarized content and class labels into a shared semantic space, FLAN-T5 capitalizes on its extensive pretraining across diverse tasks to deliver robust Zero-Shot classification. This integration not only addresses the limitations of existing ZSL methods, such as insufficient semantic representation, but also enhances the system's scalability and its ability to adapt to evolving spam patterns and adversarial tactics. The main contributions of this work are as follows:

- Proposing a novel spam detection system that integrates Zero-Shot Learning with advanced NLP techniques, using both BERT and FLAN-T5 to improve classification accuracy and adaptability.

- Utilizing BERT for noise reduction by preprocessing email content to filter out irrelevant information and generate concise summaries, enabling the system to focus on critical details for effective classification.

- Addressing key limitations in existing ZSL-based spam detection methods, particularly the lack of semantic representation and contextual understanding, which are essential for handling evolving spam content.

- Removing dependency on labeled datasets and frequent retraining, resulting in a scalable, cost-effective system that adapts to changes in spam characteristics over time and reduces operational overhead while remaining resilient to challenges such as concept drift and adversarial tactics.

The remainder of this report is structured as follows. The Literature Review examines existing spam detection techniques and highlights the limitations of traditional methods and ZSL-based approaches. The Methodology describes the proposed system and its components. The Implementation section outlines the datasets and system architecture. The Evaluation presents the results, followed by a Discussion of the findings and potential improvements. Finally, the Conclusion summarizes the contributions and suggests directions for future research.

## 2 Literature Review

This section provides the necessary background to understand the concepts and challenges related to Spam detection using machine learning.

### 2.1 Traditional Spam Detection Methods

Early approaches to spam detection relied on rule-based systems, which used predefined patterns or heuristics to identify spam messages. While these methods were simple and interpretable, they struggled to handle the growing complexity and variability of spam. To address these limitations, machine learning (ML) techniques such as Naïve Bayes, Support Vector Machines (SVMs), and Decision Trees became widely used [11, 8]. These models learned patterns from labeled training data, offering better accuracy and adaptability than rule-based systems. However, traditional ML methods have inherent limitations:

1. They require large, labeled datasets, which are expensive and time-consuming to create.

2. They are unable to adapt to "concept drift", where spam patterns evolve over time.

3. They are vulnerable to adversarial tactics, where spammers intentionally modify content to evade detection.

Deep learning models, including Convolutional Neural Networks (CNNs) and Long Short-Term Memory (LSTM) networks, have further advanced spam detection by capturing complex patterns in textual data [7]. While these models achieve high accuracy, they are computationally intensive, require extensive labeled data, and still struggle to adapt to the dynamic nature of spam.

### 2.2 The Role of Zero-Shot Learning

Zero-Shot Learning (ZSL) has emerged as a powerful machine learning paradigm, addressing key limitations of traditional supervised methods. Unlike conventional models that require labeled datasets for each task, ZSL enables the classification of unseen data by leveraging semantic relationships between input data and predefined class labels [1]. This makes it particularly effective for rapidly evolving tasks like spam detection, where new patterns and categories frequently emerge.

ZSL is especially relevant in spam detection due to its ability to adapt without relying on costly and time-intensive labeled datasets. By embedding input data and class labels into a shared semantic space, ZSL models infer relationships and classify unseen data based on contextual alignment. This scalability makes ZSL an efficient solution for handling dynamic and adversarial spam patterns.

However, traditional ZSL methods face challenges such as limited semantic understanding and difficulty managing class imbalance, common in spam detection [5]. Modern ZSL frameworks address these issues by leveraging advanced language models like FLAN-T5, which enhance contextual understanding through extensive pretraining [6]. These models enable ZSL to detect complex spam patterns and dynamically adapt to new challenges, bridging the gap between theoretical potential and practical application in spam detection.

### 2.3 Advancements in NLP and Their Impact on Spam Detection

Recent advancements in NLP have significantly enhanced the capabilities of spam detection systems. LLMs like GPT-4, BERT, and FLAN-T5 have demonstrated exceptional performance in understanding textual data by leveraging extensive pretraining on diverse tasks [13, 4]. These models excel at capturing semantic relationships and contextual details, addressing critical gaps in traditional ZSL methods. Studies in anomaly detection have shown that using models like BERT and GPT-4 for preprocessing tasks, such as summarization and feature extraction, improves the understanding and effectiveness of Zero-Shot Learning [12].

- **BERT:** As a bidirectional transformer, BERT processes text holistically, understanding the context of each word in a sentence. Its ability to generate meaningful embeddings makes it highly effective for tasks such as summarization and noise reduction in spam detection pipelines.

- **GPT-4:** As one of the most advanced transformer-based models, GPT-4 exhibits exceptional abilities in contextual understanding and text generation. Its use in preprocessing tasks, such as content summarization and feature enrichment, has proven to support better downstream Zero-Shot Learning performance [10].

By integrating ZSL with these advanced NLP models, spam detection systems can effectively address challenges such as concept drift, adversarial tactics, and the reliance on extensive labeled datasets.

## 3 Methodology

This study proposes a spam detection system that integrates Zero-Shot Learning with the help of advanced NLP models to address key challenges in spam detection. By utilizing BERT for preprocessing and FLAN-T5 for classification, the system eliminates the need for large labeled datasets, adapts to evolving spam patterns, and enhances the scalability of spam detection systems.

The methodology begins with preprocessing the raw email content to ensure it is clean, concise, and ready for classification (You can see a summary of methodology in figure 1). Initially, the raw data is cleaned by removing noise, including duplicates, stopwords, and non-alphabetic characters. This ensures that only relevant content remains, and improves the quality of the input data. Following this, the cleaned email content is processed through BERT, which generates concise summaries by distilling the most relevant information while preserving the semantic meaning. This step reduces the complexity of the input data, allowing the classification model to focus on critical details for accurate predictions.

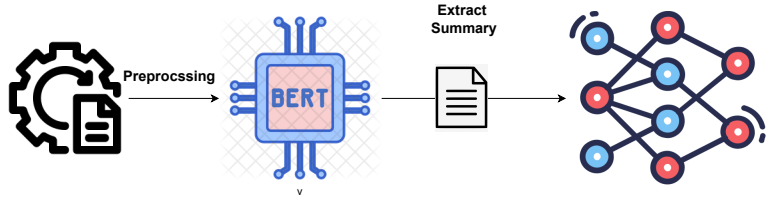

Figure 1: The Proposed Methodology Architecture

Once the email content has been summarized, it is passed to the FLAN-T5 model for classification. FLAN-T5 operates within a Zero-Shot Learning framework, embedding both the summarized content and the class labels ("spam" and "ham") into a shared semantic space. This semantic embedding allows the model to evaluate the contextual alignment between the email content and the class definitions without requiring additional labeled training data. Leveraging FLAN-T5's extensive pretraining on diverse tasks, the system achieves robust generalization, enabling it to adapt effectively to new and unseen spam patterns. By using Zero-Shot Learning, the system mitigates the challenges of concept drift and reduces the operational overhead associated with frequent retraining.

## 4 Evaluation

To evaluate the proposed approach, the Spam SMS Detection dataset was utilized. This dataset provides a realistic representation of spam and non-spam ("ham") messages, enabling the assessment of the system's performance in a practical context. The evaluation was conducted using standard metrics, including accuracy, precision, recall, and F1-score, to provide a comprehensive view of the model's effectiveness.

The results indicate a promising accuracy of 72%, alongside a micro-precision score of 0.65, a micro-recall score of 0.54, and a micro-F1 score of 0.52. These metrics demonstrate the potential of the approach in distinguishing spam messages from legitimate ones effectively. However, the relatively lower recall score highlights that the model could miss some spam messages, indicating room for improvement, particularly in detecting spam with varied and subtle patterns.

While the performance is promising, it underscores the need for further refinements to enhance the model's recall and overall robustness. Several ideas are being considered to address these limitations, such as optimizing the pre-processing stage, fine-tuning the Zero-Shot Learning process with additional data, or incorporating adversarial training to improve the system's resilience to sophisticated spam tactics.

This evaluation demonstrates the viability of the proposed method while highlighting opportunities for future enhancements to achieve even more reliable spam detection.

## 5  Discussion

This study addresses challenges such as data scarcity, concept drift, and adversarial tactics by eliminating the reliance on large labeled datasets and retraining, making it adaptable to real-world spam patterns. However, the findings highlight several limitations and open avenues for further investigation. One key limitation is the relatively low recall, which indicates that the model struggles to identify all spam messages, particularly those crafted using sophisticated or adversarial techniques. This suggests the need for more robust preprocessing and classification strategies to enhance the system's ability to detect subtle spam signals without compromising precision. Additionally, the reliance on BERT for summarization underscores the importance of ensuring that critical contextual information is preserved during preprocessing, as information loss can directly impact classification performance.

Another limitation lies in the computational overhead associated with using large language models like BERT and FLAN-T5. While these models offer significant advantages in semantic understanding and generalization, their resource-intensive nature can limit real-time deployment and scalability in resource-constrained environments. Future work should explore model optimization techniques such as distillation or pruning to reduce computational requirements while maintaining performance.

Beyond spam detection, the ZSL framework combined with LLMs presents opportunities for broader applications. For instance, extending this approach to phishing detection, fraud prevention, or misinformation classification could address similar challenges in these domains, such as data scarcity and evolving adversarial tactics. Exploring hybrid approaches that combine ZSL with Few-Shot Learning may also enhance adaptability, enabling the system to leverage minimal labeled examples in domain-specific contexts.

Finally, while this study focuses on the Spam SMS Detection dataset, evaluating the approach on a wider range of datasets, including diverse languages, domains, and spam types (e.g., phishing or business email compromise), would provide deeper insights into its generalizability. Further research could also investigate the use of alternative NLP models or multitask learning frameworks to refine the system's capabilities and broaden its applicability.

# 6 Conclusion

This study presents a novel approach to email spam detection by integrating Zero-Shot Learning with NLP techniques. The proposed system leverages BERT for summarizing email content and FLAN-T5 for Zero-Shot classification, addressing critical challenges in spam detection, including the scarcity of labeled data, evolving spam patterns, and adversarial tactics. By eliminating the dependency on extensive labeled datasets and frequent retraining, this approach offers a scalable, efficient, and future-proof solution for real-world spam detection tasks.

Using the Spam SMS Detection dataset, the system achieved an accuracy of 72%, demonstrating its effectiveness in distinguishing spam messages from legitimate ones. The findings highlight the effectiveness of combining BERT's preprocessing capabilities with FLAN-T5's adaptability in handling dynamic and unseen spam patterns. This integration improves the system's robustness to concept drift and enhances its ability to classify emails in adversarial environments. These results demonstrate the promise of ZSL and large language models in advancing spam detection systems, offering a significant step forward in addressing the limitations of traditional and deep learning-based methods.

# References

[1] N. S. Bendre, "Explainability with semantic concept composition and zero-shot learning for anomaly detection," Ph.D. dissertation, The University of Texas at San Antonio, 2021.

[2] M. Crawford, T. M. Khoshgoftaar, J. D. Prusa, A. N. Richter, and H. Al Najada, "Survey of review spam detection using machine learning techniques," *Journal of Big Data*, vol. 2, pp. 1–24, 2015.

[3] F. Fdez-Riverola, E. L. Iglesias, F. Díaz, J. R. Méndez, and J. M. Corchado, "Applying lazy learning algorithms to tackle concept drift in spam filtering," *Expert Systems with Applications*, vol. 33, no. 1, pp. 36–48, 2007.

[4] T. Koide, N. Fukushi, H. Nakano, and D. Chiba, "Chatspamdetector: Leveraging large language models for effective phishing email detection," *arXiv preprint arXiv:2402.18093*, 2024.

[5] Y. Li, R. Zhang, W. Rong, and X. Mi, "Spamdam: Towards privacy-preserving and adversary-resistant sms spam detection," *arXiv preprint arXiv:2404.09481*, 2024.

[6] H. Luo, "Prompt-learning and zero-shot text classification with domain-specific textual data," 2023.

[7] S. K. Maurya, D. Singh, and A. K. Maurya, "Deceptive opinion spam detection approaches: a literature survey," *Applied intelligence*, vol. 53, no. 2, pp. 2189–2234, 2023.

[8] S. Rao, A. K. Verma, and T. Bhatia, "A review on social spam detection: Challenges, open issues, and future directions," *Expert Systems with Applications*, vol. 186, p. 115742, 2021.

[9] S. Si, Y. Wu, L. Tang, Y. Zhang, and J. Wosik, "Evaluating the performance of chatgpt for spam email detection," *arXiv preprint arXiv:2402.15537*, 2024.

[10] J. Su, C. Jiang, X. Jin, Y. Qiao, T. Xiao, H. Ma, R. Wei, Z. Jing, J. Xu, and J. Lin, "Large language models for forecasting and anomaly detection: A systematic literature review," *arXiv preprint arXiv:2402.10350*, 2024.

[11] P. Teja Nallamothu and M. Shais Khan, "Machine learning for spam detection," *Asian Journal of Advances in Research*, vol. 6, no. 1, pp. 167–179, 2023.

[12] A. Utaliyeva, M. Pratiwi, H. Park, and Y.-H. Choi, "Chatgpt: A threat to spam filtering systems," in *2023 IEEE International Conference on High Performance Computing & Communications, Data Science & Systems, Smart City & Dependability in Sensor, Cloud & Big Data Systems & Application (HPCC/DSS/SmartCity/DependSys)*. IEEE, 2023, pp. 1043–1050.

[13] Z. Zhang, Y. Wu, H. Zhao, Z. Li, S. Zhang, X. Zhou, and X. Zhou, "Semantics-aware bert for language understanding," in *Proceedings of the AAAI conference on artificial intelligence*, vol. 34, no. 05, 2020, pp. 9628–9635.